# Strong reduction of ac losses in a superconductor strip located between superconducting ground plates

Y.A. Genenko

*Institut für Materialwissenschaft, Technische Universität Darmstadt,*
*Petersenstrasse 23, D-64287 Darmstadt, Germany*
*E-mai:* yugenen@tgm.tu-darmstadt.de

The problem of calculating the ac losses in a superconductor strip with a transport current placed inside superconducting environments is studied analytically in the frame of the critical state model. Exact results obtained by the method of images for the commonly employed flat ground plates are used to derive power losses and, consequently, the nonlinear resistance depending on the ac frequency, current amplitude and the distance to the ground plates. The resistance is strongly reduced when the distance between the strip and the shields becomes small.

Corresponding author: Tel.: +49-6151 16 4526; fax: +49-6151 16 6038.
*E-mail address:* yugenen@tgm.tu-darmstadt.de

## 1. Introduction

Shielding of superconductors may strongly change the current distribution in them coupled to the net field. Using this fact ground plates have been commonly used for homogenizing the current in superconductor strips [1-3] and weak links [4] and reducing ac losses in them. As was recently predicted special curved magnetic and superconducting surroundings may even significantly enhance the current-carrying capability of strips and weak links in the flux-free Meissner state [5-7] as well as in the flux-filled critical state [8,9]. Experimentally, critical current enhancements [10,11] and ac losses reduction [12-14] were observed recently in magnetically shielded strips and filaments.

Using critical state model [15] and recent exact results on the current distributions in shielded strips [9] we calculate in this work power losses arising in a superconductor strip carrying ac transport current and located between two loss-free superconducting ground plates, a nearly realistic geometry shortly studied numerically [16]. The calculated current-voltage characteristics are strongly nonlinear because of hysteretic magnetic behaviour in the critical state while the resistance exhibits substantial reduction with decreasing of distance between the shields and the strip.

## 2. Current distribution in the shielded strip

Let us consider a strip of thickness $d$ that is much less than the strip width, $W$. The magnetic field around the strip may be expressed, to an accuracy of $d/W<<1$, through the sheet current $J$ which is the current density integrated over the strip thickness [17]. To this end, the strip may be considered as one of zero thickness. Let the strip occupy the region $(-W/2, W/2)$ on the $x$-axis and be infinite in the current direction along the $z$-axis.

We consider the shielding configuration consisted of two superconductor ground plates of thickness much larger than $\lambda$, the London penetration depth, located parallel to the strip so that their inward boundaries take positions $y = \pm a$. The ground plates will be treated macroscopically as if the London penetration depth $\lambda = 0$ therefore magnetic field lines must be tangential to the shield surfaces. According to this assumption the ground plates will not contribute to the ac losses in the system.

For description of the magnetic field penetration in the strip we adopt the Bean model of the critical state [15] adapted by Norris to the strip geometry [18]. This assumes that magnetic flux lines start to move when the local current density reaches a critical value $j_c$, determined by the pinning of magnetic vortices by materials inhomogeneities. The critical state of the strip partly flux-filled from the edges may be expressed by equations [18]

$$\begin{cases} H(x) = 0, & |x| < b/2 < W/2 \\ J(x) = J_c, & b/2 < |x| < W/2 \end{cases} \quad (1)$$

with $b$ is the width of the flux-free zone, $J_c = j_c d$ and $H(x)$ means the *y*-component of the magnetic field perpendicular to the strip plane $y = 0$ and taken at $y = 0$.

Using method of images the magnetic field $\mathbf{H}$ in between the shields may be represented as the field produced by the strip itself superimposed by the fields of strip images which results in the expression

$$H(x) = \frac{1}{4a} \int_{-W/2}^{W/2} du \, \frac{J(u)}{\sinh(\pi(x-u)/2a)} \, . \quad (2)$$

The current distribution in the critical state of the shielded strip satisfying (1) and (2) reads [9]

$$J(x) = \begin{cases} \dfrac{2J_c}{\pi} \arctan \sqrt{\dfrac{(l^2 - p^2)(1-\sigma^2)}{(1-l^2)(p^2 - \sigma^2)}}, & |x| < b/2 < W/2 \\ J_c, & b/2 < |x| < W/2 \end{cases} \quad (3)$$

with $\sigma = \tanh(\pi x/2a), l = \tanh(\pi W/4a), p = \tanh(\pi b/4a)$.

Current profiles for different distances between the strip and the shields *a* are shown in Fig. 1. It is seen that for $a << W$ the current is practically constant in the flux free zone too. This enables a simple approximation for the total current $I$ with an inaccuracy not more than 1% for $a < 0.1$:

$$I = I_c \left\{ 1 - \frac{b}{\pi W} \arccos \left[ 1 - 2 \exp\left( -\frac{\pi(W-b)}{2a} \right) \right] \right\} \quad (4)$$

where the critical current of the strip $I_c = J_c W$. For almost complete flux penetration $b << W$ and $I \to I_c$ as

$$I = I_c \left[ 1 - \frac{2b}{\pi W} \exp\left( \frac{-\pi W}{4a} \right) \right] \qquad (5)$$

while for weak penetration $W - b \ll a < W$ and the total current

$$I = I_c \sqrt{\frac{2(W-b)}{\pi a}} \quad . \qquad (6)$$

The corresponding magnetic field may be found by substitution of Eq.(3) in Eq.(2) and consequent integration.

### 3. ac losses in the shielded strip

Evolution of magnetic field profiles in the critical state of the superconductor strip when the ac transport current $I_{tr}(t) = I_0 \cos(\omega t)$ is applied were studied in details in Refs. [18,19]. It was shown that power losses per unit length of the strip $P(\omega) = (\omega/\pi) U_{HC}$ where the energy $U_{HC}$ dissipated in the strip during the half cycle is uniquely determined by the field profile at the peak current $I_{tr}(t) = I_0$:

$$U_{HC} = 2\mu_0 J_c \int_{b/2}^{W/2} dx \int_{b/2}^{x} du \, H(u) \qquad (7)$$

Taking into account that the current is almost constant in the flux-free zone when $a \ll W$ one obtains in this case approximately

$$H(u) = \frac{J_c}{2\pi} \left\{ \ln \left[ \frac{\tanh\frac{\pi}{4a}\left(u+\frac{w}{2}\right) \tanh\frac{\pi}{4a}\left(u-\frac{b}{2}\right)}{\tanh\frac{\pi}{4a}\left(u+\frac{b}{2}\right) \tanh\frac{\pi}{4a}\left(\frac{w}{2}-u\right)} \right] + \frac{2}{\pi} \arctan\sqrt{\frac{l^2-p^2}{1-l^2}} \ln \left[ \frac{\tanh\frac{\pi}{4a}\left(u+\frac{b}{2}\right)}{\tanh\frac{\pi}{4a}\left(u-\frac{b}{2}\right)} \right] \right\}. \quad (8)$$

The difference between the field patterns in shielded and unshielded strips for maximum flux penetration at the same value of the transport current amplitude is presented in Fig. 2. Reduced flux-filled margins and depressed field amplitude in the shielded strip (Fig. 2b) in comparison with the unshielded one (Fig. 2a) leads to the reduction of ac losses in the former.

Defining the resistance per unit length of the strip as $R = 2P/I_0^2$ one finds for complete penetration at $b = 0$ and $I_0 = I_c$

$$R = \mu_0 \omega \frac{16A}{\pi^4} \left(\frac{a}{W}\right)^2 \qquad (9)$$

with $A = 1.054$. For weak penetration $W - b \ll a < W$ and $I_0 \ll I_c$

$$R = \mu_0 \omega \frac{1}{8} \left(\frac{a}{W}\right)^2 \left(\frac{I_0}{I_c}\right)^2 \ln\left(\frac{4W I_c^2}{\pi a I_0^2}\right) \qquad (10)$$

where we have used Eq. (6). The current-voltage characteristics computed for $I_0 < I_c$ with the help of approximation (8) are shown in Fig. 2.

## 4. Conclusions

Type II superconductor strips with strong pinning are known to exhibit intrinsic nonlinear impedance because of hysteretic magnetization which may be understood using the critical state model [16,18,20]. The resulting large resistance at high frequencies handicaps implementation of microstrips in microwave devices. The strong reduction of the strip resistance by superconducting ground plates is conventionally used to improve the device performance but a proper understanding of ac losses in shielded strips and in the shields is yet to be achieved. In this work we obtained analytical expressions for the shielded strip resistance which scales down with the decreasing distance to the shields $a$ roughly as $(a/W)^2$ in agreement with the computations in Ref. [16]. The reason for this strong reduction is the decrease of field peaks at the strip edges as well as the fact that the major part of the strip remains practically flux-free up to the current amplitudes close to the critical current of the strip.

Discussions with Prof. H. Rauh are gratefully acknowledged.

**Figure captions**

**Fig. 1.** Sheet current distributions in the shielded strip for the same value of the total transport current $I = 0.9 I_c$ but for the different values of the spacing $a$ between the strip and the ground plates.

**Fig. 2.** Magnetic field profiles in the unshielded (a) and shielded (b) strips at the moment of maximum flux penetration for different values of the transport current amplitude $I$. The spacing between the strip and the ground plates $a = 0.1 W$

**Fig. 3.** ac voltage per unit length of a shielded strip against the amplitude of ac transport current for small values of the spacing $a$ between the strip and the ground plates.

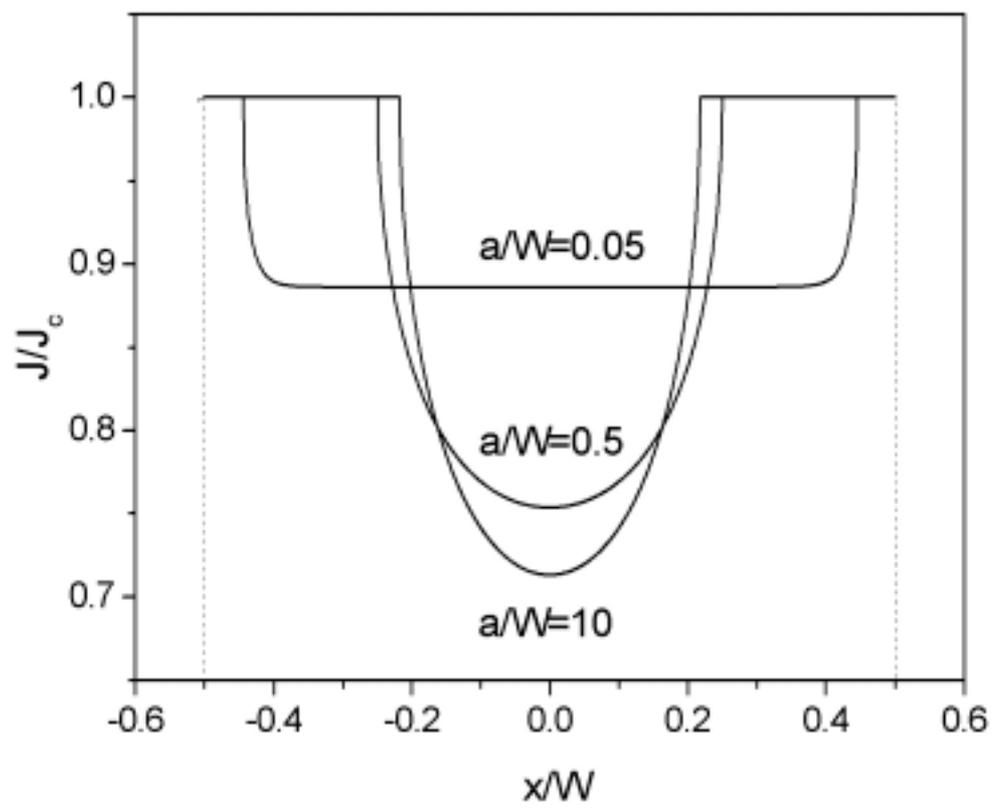

Fig. 1

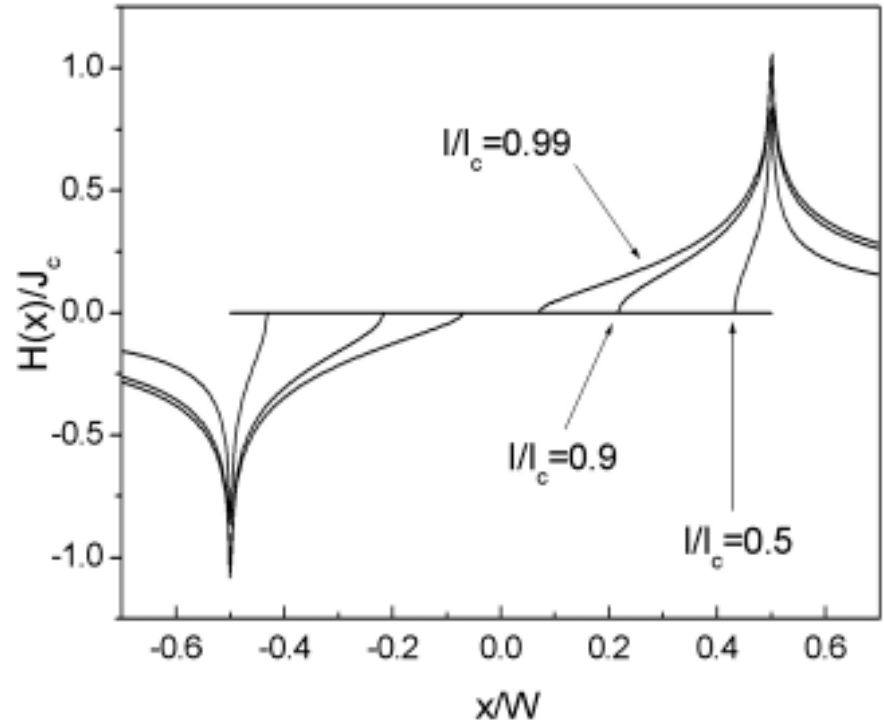

(a)

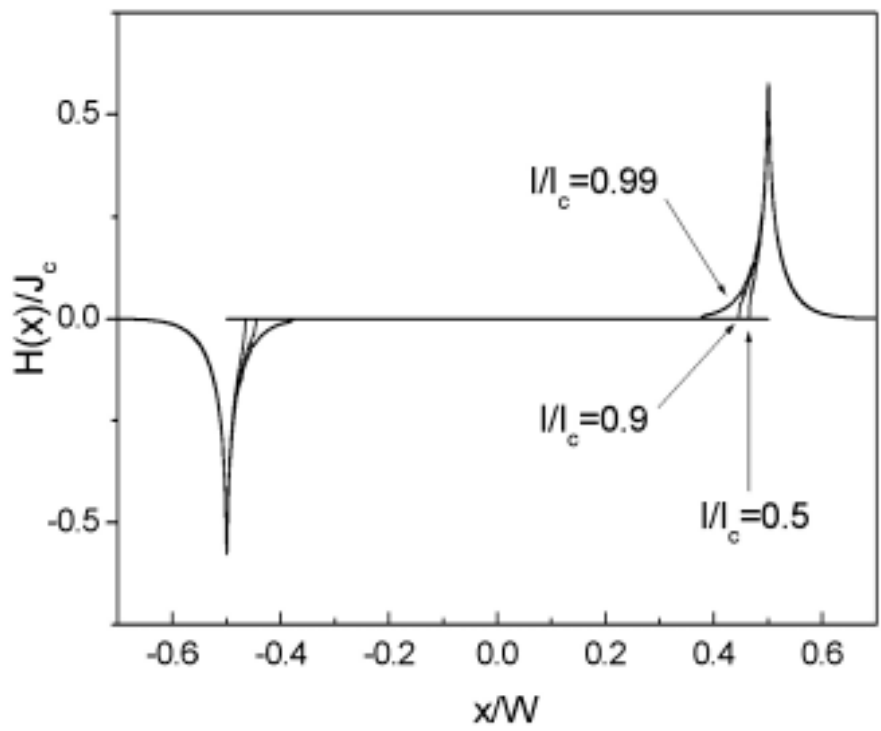

(b)

Fig. 2

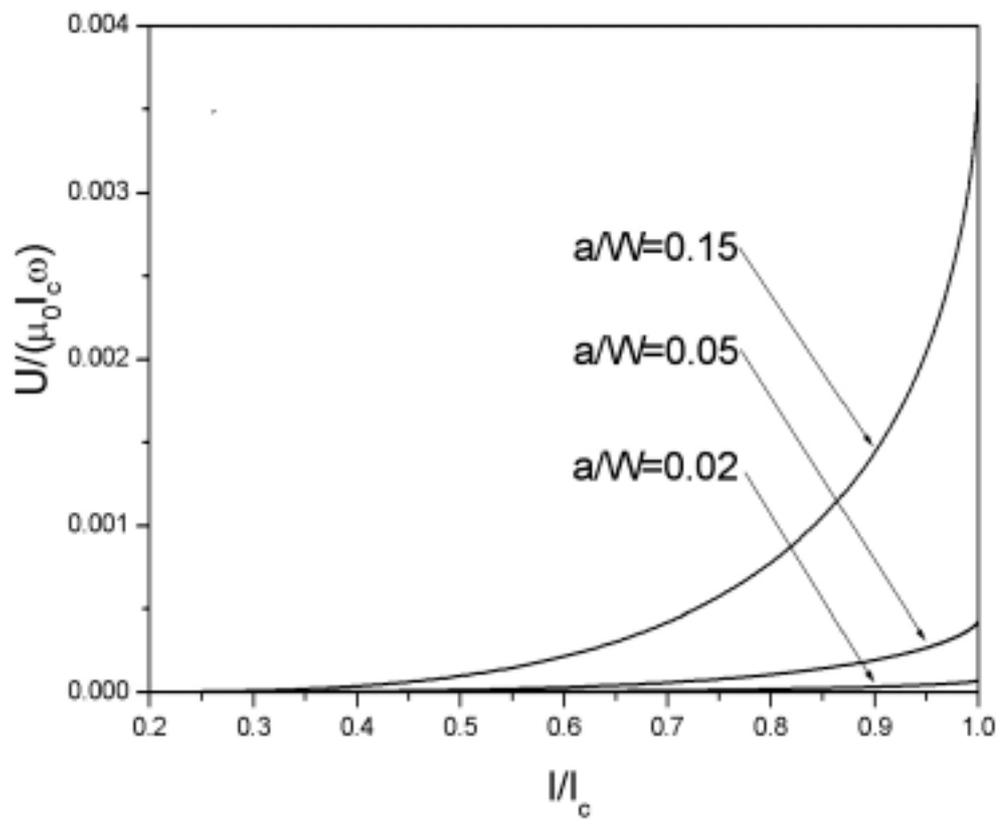

Fig. 3